\documentclass{article}
\usepackage{spconf,amsmath,graphicx}

\input{mysymbol.sty}

\usepackage[utf8]{inputenc} % allow utf-8 input
\usepackage[T1]{fontenc}    % use 8-bit T1 fonts
\usepackage{hyperref}       % hyperlinks
\usepackage{url}            % simple URL typesetting
\usepackage{booktabs}       % professional-quality tables
\usepackage{amsfonts}       % blackboard math symbols
\usepackage{nicefrac}       % compact symbols for 1/2, etc.
\usepackage{microtype}      % microtypography
\usepackage{lipsum}
\usepackage{amsmath}
\usepackage{amssymb}        % as for the triangleq for definition of symbols
\usepackage{amsthm} % for theorem and definitions
\theoremstyle{plain}
\theoremstyle{definition}
\usepackage{verbatim}
\usepackage{empheq,float}
\usepackage{graphicx,caption,subcaption,balance}
\usepackage{stfloats}
\usepackage[dvipsnames]{xcolor}

\usepackage{epstopdf}
\usepackage{algorithm}
\usepackage{algpseudocode}
\usepackage{mathrsfs}
\usepackage{psfrag}

\title{Online Time-Varying Topology Identification via Prediction-Correction Algorithms}
\name{Alberto Natali, Mario Coutino, Elvin Isufi and Geert Leus\thanks{E-mails: \{a.natali; m.a.coutinominguez; e.isufi-1; g.j.t.leus\}@tudelft.nl; This work was sponsored in part by Theory Lab, Central Research Institute, 2012 Labs, Huawei Technologies Co.,Ltd, and  by the KAUST-MIT-TUD consortium grant OSR-2015-Sensors-2700. Mario Coutino is partially supported by CONACYT.}}
\address{Faculty of Electrical Engineering, Mathematics and Computer Science\\ Delft University of Technology, Delft, The Netherlands}
\begin{document}
\ninept
\maketitle
\begin{abstract}
 Signal processing and machine learning algorithms for data supported over graphs, require the knowledge of the graph topology. Unless this information is given by the physics of the problem (e.g., water supply networks, power grids), the topology has to be learned from data. Topology identification is a challenging task, as the problem is often ill-posed, and becomes even harder when the graph structure is time-varying. In this paper, we address the problem of dynamic topology identification by building on recent results from time-varying optimization, devising a general-purpose online algorithm operating in non-stationary environments. Because of its iteration-constrained nature, the proposed approach exhibits an intrinsic temporal-regularization of the graph topology without explicitly enforcing it. As a case-study, we specialize our method to the Gaussian graphical model (GGM) problem and corroborate its performance.

\end{abstract}
\section{Introduction}
\label{sec:intro}

The knowledge of the graph topology is a \emph{requirement} for processing tools operating on data residing on top of networks, often conceptualized as graph signals~\cite{shuman2013emerging,6409473}. Unless the topology is known beforehand like in water supply networks or power grids~\cite{newman2018networks}, it has to be inferred from the available networked data. 
This challenging task is known in literature as \textit{network topology inference} or \textit{graph
learning}~\cite{mateos2019connecting,dong2019learning}, and it becomes even harder when the  graph topology is dynamic, i.e., it changes over time, like in brain functional connectivity~\cite{preti2017dynamic} and biological networks~\cite{kim2014inference}.

When the network is dynamic, the dependence of the graph structure is reflected on the underlying time-varying distribution of the observable data. Thus, the task of dynamic network topology identification is to use the stream of data to infer the evolution of the network structure in an online fashion. This online inference task allows a network invigilator to promptly detect anomalies, e.g., uncommon financial transactions among users, or to perform effective decision making, e.g., load redistribution in smart grids.%\footnote{In this discussion, we do not specify whether is the change of data distribution that causes a change in the topology or the opposite.}.

Previous works in this area, e.g.~\cite{kalofolias2017learningtvgraphs,yamada2020time}, learn a sequence of graphs by enforcing a prior (smoothness or sparsity) on the edges of consecutive graphs. Adopting a similar approach, the work in~\cite{hallac2017network}  extends the graphical Lasso~\cite{friedman2008sparse} to  estimate the dynamic topology using the  alternating direction method of multipliers (ADMM)~\cite{boyd2011distributed}. In addition to these works, the inference of causal relationships in the network structure, i.e., directed edges, has been considered in~\cite{baingana2017cascades}. For a complete review of causal dynamic topology inference, the reader is referred to~\cite{giannakis2018topology}.

The aspect that brings all previous approaches together is the following same \emph{processing chain}: $i)$ data-collection; $ii)$ data-splitting (windowing); $iii)$ batch-processing with structural constraints between consecutive graph topologies. This processing chain fails to address the \textit{online (data-streaming) setting}, recently investigated in~\cite{vlaski2018online}, which assumes graph data generated by a heat diffusion process, and by~\cite{shafipour2020online}, which assumes data to be graph stationarity, i.e., the covariance matrix of the data and the matrix representation of the network commute. Differently from these works, our framework is more general as it does not require such assumptions.

To address the dynamic network topology inference problem, we develop an adaptive algorithm based on the time-varying convex optimization framework~\cite{simonetto2016class,simonetto2020time}. This framework operates on-the-fly, thus adapting to non-stationary dynamics, and runs without requiring the entire sequence of observations. The devised adaptive algorithm does not assume knowledge of the time instants at which the topology changes and it implicitly applies a temporal regularization due to its early-stopping behavior, i.e., limited iteration budget. We demonstrate the validity of our approach by focusing on the well-known Gaussian graphical model (GGM) problem. Although the GGM problem does not consider the causal scenario, i.e., non-symmetric network topology, we stress that the proposed dynamic network topology tracking framework  can be applied to this scenario as well.
    
%%%%%%%%%%%%%%%%%%%%%%%%%%%%%%%%%%%%%%%%%%%%%%%%%

\section{Preliminaries}
\vspace{-3mm}
\label{sec:preliminaries}
\subsection{Graphs and Signals over Graphs}
 Consider a scenario in which the data of interest reside in a non-Euclidean domain described by the undirected graph $\ccalG~=~\{\ccalV, \ccalE, \bbS\}$, where $\ccalV=\{1, \ldots, N\}$ and $\ccalE \subseteq \mathcal{V} \times \mathcal{V}$ are the vertex and edge set, respectively; and $\bbS$ is an $N \times N$ symmetric matrix that represents the graph structure, which entries $\left[\bbS\right]_{i,j}$ are nonzero only if $(j,i) \in \ccalE$,  for $i\neq j$. Matrix $\bbS$ is typically called the graph shift operator (GSO)~\cite{6409473}, and examples include the (weighted) adjacency matrix $\bbW$~\cite{6409473} and the graph Laplacian $\bbL$~\cite{shuman2013emerging}. By associating to each node $i \in \ccalV$ a scalar value $x_i$, we can define $\bbx \in \mathbb{R}^{N}$ as a \emph{graph signal} mapping the node set to the set of real vectors.
 
 \subsection{Topology Identification}
\textbf{Static.} When the knowledge of the topology of the network is not available, we encounter the problem of learning the network structure from data, a problem known as \emph{topology identification or graph learning}.
Formally, consider the matrix $\bbX=[\bbx_1, \ldots, \bbx_T]$ that stacks $T$ graph signals of dimension $N$ arising from an unknown graph $\ccalG=(\ccalV,\ccalE, \bbS)$. The goal is then to infer the \textit{latent} underlying topology of the graph by these data, which amounts to estimating the GSO. In this work, we focus particularly on the Gaussian graphical model (GGM) problem, which we recall, to pave the way for the illustration of our (dynamic) network inference framework.

\smallskip
\noindent\textbf{Gaussian Graphical Model.} Assume each graph signal $\bbx$ to be drawn from a multivariate Gaussian distribution $\ccalN(\boldsymbol{\mu}, \mathbf{\Sigma})$ with mean $\boldsymbol{\mu}$ and covariance matrix $\mathbf{\Sigma} \in \mathbb{S}_{++}^{N}$. By denoting with $\bbS=\mathbf{\Sigma}^{-1}$ the \emph{precision} matrix, \emph{graph learning in a GGM amounts to precision matrix estimation}. This is because a missing edge in $\ccalG$ corresponds to a zero entry in the precision matrix, and  consequently to the conditional independence of the related  entities of the graph. Formally, the maximum likelihood estimation (MLE) for $\bbS$ reads as:
\begin{equation}
\label{eq:ggm}
    \underset{\bbS \in   \mathbb{S}_{++}^{N}}{\operatorname{minimize}} -\log \operatorname{det}(\bbS) +\operatorname{tr}(\bbS \hat{\mathbf{\Sigma}}),
\end{equation}
where $\hat{\mathbf{\Sigma}}$ is the sample covariance matrix and $\mathbb{S}_{++}^{N}$ is the set of $N \times N$ positive definite matrices.

\smallskip
\noindent\textbf{Dynamic. }Our focus is, however, on scenarios where the graph topology changes over time, e.g., social interaction or brain connections. These changing interactions are represented by a time-varying topology, which we express by  the sequence of graphs $\{\ccalG_{t}= (\ccalV, \ccalE_{t}, \bbS_{t})\}_{t=1}^{\infty}$ with associated GSOs $\{\bbS_{t}\}_{t=1}^\infty$. This formulation can be seen as the sampling of a time-continuous dynamic topology using a sampling period of $h$.

We consider time-varying graph signals $\{ {\bf x}_t \}_{t=1}^\infty$ arising from an unknown time-varying graph $\{\ccalG_{t}\}_{t=1}^\infty$. Akin to the static case, now \textit{the goal is to infer the underlying time-varying topology of the graph from $\{ {\bf x}_t \}_{t=1}^\infty$}, which is tantamount to the estimation of the matrix sequence $\{\bbS_{t}\}_{t=1}^{\infty}$.

Using the above formalism, most of the current works have relied on batch approaches to tackle the dynamic topology inference setting. That is, after all possible data have been collected, pre-processed, and then split into (possibly overlapping) temporally-contiguous windows, each window is associated with a graph topology. This topology has been further constrained to be \textit{close} to those of temporally-adjacent windows, by solving a regularized version of~\eqref{eq:ggm}, see, e.g.,~\cite{hallac2017network}. These approaches $i)$ require all windows at once to form the dynamic topology estimate and $ii)$ need to select the time instances where to split the data into windows, i.e., the approximate time instances where the topology changes. These form a limitation because in many scenarios the memory storage is an issue, batch processing is computationally expensive, or a control agent may need to take real-time decisions.

To alleviate these issues, we devise an on-the-fly adaptive algorithm, using time-varying optimization~\cite{simonetto2020time}, that updates the solution as samples come into the system, thus avoiding assumptions on the time instants at which the topology changes. Because of its iteration-budget constraints, the adaptive method exhibits intrinsic regularization properties without the need of explicit regularizers. This is advantageous since we do not need to rely on a specific regularization assumption such as smoothness or sparsity, which may not hold in practice. 

In Section~\ref{sec:tvti}, we introduce the time-varying optimization framework for the dynamic graph learning problem and, in Section~\ref{sec:tvggm}, we specialize it to the GGM problem.

%%%%%%%%%%%%%%%%%%%%%%%%%%%%
\vspace{-2mm}
\section{Time-Varying Topology Identification}
\label{sec:tvti}
Conventional optimization methods can no longer be taken for granted when dealing with dynamic systems whose internal parameters are continuously evolving and for which fast decision-making tasks are required. Fortunately, advances have been made in the field of time-varying optimization~\cite{simonetto2016class,simonetto2020time}. We leverage these ideas and propose a framework that given a time-varying cost function $f(\bbS,t)$ maintains an estimate solution $\hat{\bbS}_{t}$, at time instant $t$, of the time-varying optimization problem $\min_{\bbS\in\ccalS}f(\bbS,t)$ for some feasible set $\ccalS$, and then \emph{predict} how this solution evolves in the next time step $t+1$. This prediction is then \emph{corrected} after a new datum is available at time $t+1$.

Formally, we aim to solve the sequence of time-invariant problems of the form:
\begin{equation}
\label{eq:main-problem}
    \bbS_{t}^{*}:=\underset{\bbS \in \ccalS}{\operatorname{argmin}} f(\bbS ; t), \quad  t \in \mathbb{N}
\end{equation}
where $\ccalS$ is a convex set defining the feasible set of GSOs\footnote{Note that the temporal variability of the function is due to the graph signals acquired up to the time instant of interest.}. As the process is sampled at given intervals and~\eqref{eq:main-problem} is typically solved by iterative methods, in most instances, the sampling period will only allow for few iterations of the selected solver, (possibly) leading to a suboptimal solution. Due to its practical relevance, this is the setting of our interest.

It can be shown that solving problem \eqref{eq:main-problem}, at time $t$, is equivalent to solving the generalized equation:

\begin{equation}
\label{eq:generalized-equation}
    \bbR(t):=\nabla_{\bbS} f\left(\bbS_{t}^{*}; t\right) + \ccalN_{\ccalS}\left(\bbS_t^{*}\right) \ni \mathbf{0},
\end{equation}
where $\nabla_{\bbS} f(\cdot) \in \mathbb{R}^{N}$ is the gradient of $f$ with respect to $\bbS$ and $\ccalN_{\ccalS}(\cdot) : \mathbb{R}^{N \times N} \xrightarrow{} \mathbb{R}^{N \times N}$ is the normal cone operator (the sub-differential of the indicator function).

\subsection{Prediction}
Suppose now that $\hat{\bbS}_t$ is an approximate solution for \eqref{eq:generalized-equation} at time $t$, i.e. a point estimate for which the residual $\bbR(t) \approx 0$. The prediction step seeks an approximate optimizer for \eqref{eq:main-problem} at $t+1$, with only the information available up to $t$.

Denoting by $\hat{\bbS}_{t+1|t}$ the output of the prediction step, the goal is then to solve:
\begin{align}
\nabla_{\bbS} f\left(\hat{\bbS}_{t+1|t}; t+1\right) + \ccalN_{S}\left(\hat{\bbS}_{t+1|t}; t+1\right) \ni \mathbf{0}
\end{align}
which is impossible at time $t$. However, we can perturb the above generalized equation, by means of backward Taylor expansion, and obtain the generalized system:
\begin{align}
    \label{eq:perturbed-equation}
    \nabla_{\bbS} f(\hat{\bbS}_t ; t) 
&+ \nabla_{\bbS \bbS} f(\hat{\bbS}_t ; t) : (\hat{\bbS}_{t+1|t} - \hat{\bbS}_t) \nonumber\\ &+ \nabla_{t \bbS} f(\hat{\bbS}_t ; t) h + \ccalN_{S}\left(\hat{\bbS}_{t+1|t}\right) \ni \mathbf{0},
\end{align}
where $\nabla_{\bbS \bbS} f (\cdot) \in \mathbb{R}^{N \times N \times N \times N}$ is the Hessian tensor of $f$ with respect to $\bbS$, $\nabla_{t \bbS} f(\cdot) \in \mathbb{R}^{N \times N}$ is the partial derivative of the gradient of $f$ w.r.t.
the time $t$, and $:$ denotes the (scalar) inner product between arguments.
% the $4$th order Hessian tensor of the function and the differential. 
We can notice that the formulation in~\eqref{eq:perturbed-equation} is equivalent to the constrained optimization problem~\eqref{eq:problem-perturbed-equation} at the bottom of the next page, where the normal cone operator leads to the definition of the constraint set over which the computation is performed. Intuitively, setting to zero the gradient of the cost function in \eqref{eq:problem-perturbed-equation}, and taking into account its feasible set, leads to \eqref{eq:perturbed-equation}.
\begin{figure*}[b!]
%\hline
\centering
\begin{equation}
    \label{eq:problem-perturbed-equation}
\hat{\bbS}_{t+1 \mid t}=\underset{\bbS \in \ccalS}{\operatorname{argmin}}\left\{\frac{1}{2} \bbS : \nabla_{\bbS\bbS} f\left(\hat{\bbS}_{t} ; t\right): \bbS
 +\left[\nabla_{\bbS} f\left(\hat{\bbS}_{t} ; t\right)+h \nabla_{t \bbS} f\left(\hat{\bbS}_{t} ; t\right)-\nabla_{\bbS \bbS} f\left(\hat{\bbS}_{t} ; t\right) :\hat{\bbS}_{t}\right]:\bbS\right\}
\end{equation}
\vspace{-2cm}
\end{figure*}

Although~\eqref{eq:problem-perturbed-equation} is a constrained quadratic optimization problem, and its solution can be efficiently found, its exact solution might incur high computational costs. Therefore, when solving~\eqref{eq:problem-perturbed-equation} we approximate its solution by means of a few projected gradient (PG) descent steps. So, letting $\hat{\bbS}^{0}$ be a dummy variable initialized as $\hat{\bbS}^{0}= \hat{\bbS}_t$, we perform the following steps:
\begin{align}
\label{eq:prediction}
    \hat{\bbS}^{p+1}
&=  \mathbb{P}_{\ccalS}[ \hat{\bbS}^{p} \nonumber - \alpha (\nabla_{\bbS} f(\hat{\bbS}_t ; t) 
+\nabla_{\bbS \bbS} f(\hat{\bbS}_t ; t) :(\hat{\bbS}^{p}-\hat{\bbS}_{t}) \nonumber \\ 
& \;\;\;\;+ \nabla_{t \bbS} f(\hat{\bbS}_t ; t) h) ]
\end{align}
for $p=0,1, \ldots, P-1$, where $P$ is a predefined number of PG steps, $\alpha > 0$ is the stepsize, and $\mathbb{P}_{\ccalS}$ it the projection operator over the convex set $\ccalS$. Once the $P$ steps are performed, the prediction $\hat{\bbS}_{t+1 \mid t}$ is set to:
\begin{equation}
    \hat{\bbS}_{t+1 \mid t}= \hat{\bbS}^{P}
\end{equation}
which is expected to approximate the optimal solution $\bbS_{t+1}^{*}$ at time $t+1$.

\subsection{Correction}

When new graph signals are acquired, i.e. the new cost function $f(\cdot; t+1)$ becomes available, the correction step refines the estimate of the optimal solution $\hat{\bbS}_{t+1 \mid t}$. To do so, we also perform a set of PG steps. In particular, by setting to $\hat{\bbS}^{0}= \hat{\bbS}_{t+1 \mid t}$, we consider the following PG steps:
\begin{equation}
\label{eq:correction}
     \hat{\bbS}^{c+1}
=\mathbb{P}_{\ccalS} \left[ \hat{\bbS}^{c} - \beta \nabla_{\bbS} f(\hat{\bbS}^{c}; t+1)  \right],
\end{equation}
for $c=0,1, \ldots, C-1$. The refined estimate of the optimal solution $\bbS_{t+1}^{*}$ is then set as $\hat{\bbS}_{t+1}= \hat{\bbS}^{C}$. Notice that we can also use a (quasi)-Newton update as an alternative to the PG step in \eqref{eq:correction} if we can afford the related computational cost.

The general time-varying optimization framework, adapted to our case, is summarized in Algorithm 1.
\begin{algorithm}[t]
\begin{algorithmic}[1]
\Require Feasible $\hat{\bbS}_0$, $f(\bbS; t_0)$, $P$, $C$
\For{$t=0,1, \ldots$}
\State // \textit{Prediction}
\State Initialize $\hat{\bbS}^0= \hat{\bbS}_t$

\For{$p= 0,1, \ldots, P-1$}

    Predict $\hat{\bbS}^{p+1}$ with \eqref{eq:prediction}

\EndFor

Set the predicted variable $ \hat{\bbS}_{t+1 \mid t}= \hat{\bbS}^{P}.
$
\State \textit{time $t+1$: new data arrive}

\State // \textit{Correction}

\State Initialize the corrected variable $\hat{\bbS}^0= \hat{\bbS}_{t+1 \mid t}$

\For{$c= 0,1, \ldots, C-1$}

    Predict $\hat{\bbS}^{c+1}$ with \eqref{eq:correction}

\EndFor

Set the corrected variable $\hat{\bbS}_{t+1}= \hat{\bbS}^{C}$
\EndFor
\end{algorithmic}
\caption{Time-Varying Topology Identification}
\label{alg:complete}
\end{algorithm}

\section{Time-Varying Gaussian Graphical Models}
\label{sec:tvggm}
The previous problem formulation has been developed without a particular specification of the cost function $f(\cdot;t)$. This enables us to extend a large class of ``static'' topology identification algorithms. That is, given a cost function, the framework only requires access to its first- and second-oder derivatives. In the following, we specialize this framework to an online GGM setting.

Denote with ${\mathcal X}_t = \{ {\bf x}_1, \dots, {\bf x}_t \}$ the set of graph signals available up to time $t$. Let also $\hat{\mathbf{\Sigma}}_t$ be the empirical covariance matrix computed with the signals up to time $t$ in a weighted moving average fashion, i.e., $\hat{\mathbf{\Sigma}}_{t}= \gamma \hat{\mathbf{\Sigma}}_{t-1} + (1-\gamma) \bbx_{t}\bbx_{t}^\top$, where $\gamma \in (0,1)$ plays the role of forgetting factor, required in non-stationary environments. Then the GGM problem, adopted to a time-varying setting, reads as:
\begin{equation}
\label{eq:graphical-lasso}
    \bbS^*_t:= \underset{\bbS \in \mathcal{S}}{\operatorname{argmin}} \{ f(\bbS; t) := -\log \operatorname{det}(\bbS) +\operatorname{tr}(\bbS \hat{\mathbf{\Sigma}}_{t})\},
\end{equation}
where $\bbS$ is the precision matrix, and $\ccalS$ defines the (convex) constraint set of valid solutions for the problem, in this case coinciding with the set of positive definite matrices. For other commonly-used sets for GSOs and/or graph matrices see, e.g.,~\cite{segarra2017network}. %cite segarra

While problem~\eqref{eq:graphical-lasso} can be solved at each time instant from scratch, this induces a large computational overhead. Thus, an approach which \emph{tracks} the optimizer $\bbS_t^*$ based on previous estimates, and the   available graph signals, is highly attractive. 

\subsection{Implementation}
To apply the presented time-varying topology identification method to the GGM problem, we need to compute the expressions of the gradients $\nabla_{\bbS } f\left(\bbS  ; t \right)$, $\nabla_{t \bbS} f(\bbS ; t)$, and the Hessian $\nabla_{\bbS \bbS} f\left(\bbS ; t\right)$ for the prediction step. Since $\bbS$ is symmetric, we can reduce the number of independent variables from $N^2$ to $N(N+1)/2$. The latter is enforced by means of representing 
%To maintain the algorithmic description more understandable, we will work in the half-vec space, i.e. we will represent 
the matrix $\bbS$ with its half-vectorization form, i.e., $\bbs=\operatorname{vech}(\bbS) \in \mathbb{R}^{N(N+1)/2}$. We further introduce the elimination matrix $\bbE \in \mathbb{R}^{[N(N+1)/2] \times N}$ and the duplication matrix $\bbD \in \mathbb{R}^{N \times [N(N+1)/2]}$ which selects the unique entries of $\bbS$, i.e., $\bbE\operatorname{vec}(\bbS) =\bbs$, and duplicates the entries of $\bbs$, i.e, $\bbD\bbs = \operatorname{vec}(\bbS)$, respectively. 

The gradient and the Hessian of the function $f$ in the half-vectorization space can be derived as: 
\vspace{-0.5mm}
\begin{align*}
\bbh(\bbS;t) &= \frac{\partial f(\bbS ; t)}{ \partial \bbs}=  \bbD^{\top} \operatorname{vec} (\hat{\mathbf{\Sigma}}_t - \bbS^{-1}) \\ \\
\bbH(\bbS)& = \frac{\partial^{2} f(\bbS;t)}{ \partial^{2}  \bbs} = \bbD^{\top} (\bbS \otimes \bbS )^{-1} \bbD,
\end{align*}
respectively. Because we are in a discrete-time setting, the time derivative of the gradient is given by the partial mixed-order derivative:
$$
\bbg_t=\bbD^{\top} \operatorname{vec} (\hat{\mathbf{\Sigma}}_{t} - \hat{\mathbf{\Sigma}}_{t-1}).
$$
Now, by defining $\hat{\bbs}_t:= \operatorname{vech(\hat{\bbS}_t)} \in \mathbb{R}^{N(N+1)/2}$, $\bbh_{t}~:=~\bbh(\hat{\bbS}_t;t)$ and $ \bbH_t~: =~\bbH(\hat{\bbS}_t)$, we have that:
\begin{itemize}
    \item \textbf{Prediction:} with $\hat{\bbs}^{0}$ a dummy variable initialized as $\hat{\bbs}^{0}= \hat{\bbs}_t$:
        \begin{equation}
            \hat{\bbs}^{p+1}
            =\mathbb{P}_{\ccalS}\left[\hat{\bbs}^{p} - 2\alpha \left(\bbh_t 
            + \bbH_t\left(\hat{\bbs}^{p}-\hat{\bbs}_{t}\right)  + h\bbg_t  \right)  \right]
        \end{equation}
        for $p=0,1, \ldots, P-1$. Then, the prediction $\hat{\bbs}_{t+1 \mid t}$ is set to $\hat{\bbs}_{t+1 \mid t}= \hat{\bbs}^{P}$.

        \item \textbf{Correction}:  by setting  $\hat{\bbs}^{0}= \hat{\bbs}_{t+1 \mid t}$:
            \begin{equation}
                 \hat{\bbs}^{c+1}
            =\mathbb{P}_{\ccalS} \left[ \hat{\bbs}^{c} - \beta \bbh_{t+1}  \right],
            \end{equation}
            for $c=0,1, \ldots, C-1$
            %. The estimate of the optimal solution $\bbs^{*}(t_{k+1})$ is then computed 
            and $\hat{\bbs}_{t+1}= \hat{\bbs}^{C}$.
    \end{itemize}

\section{Numerical Results}

\begin{figure*}%
\centering
\begin{subfigure}{0.33\textwidth}
\centering
%\psfrag{NMSE}[bc]{\footnotesize NMSE}
\includegraphics[width=.95\textwidth]{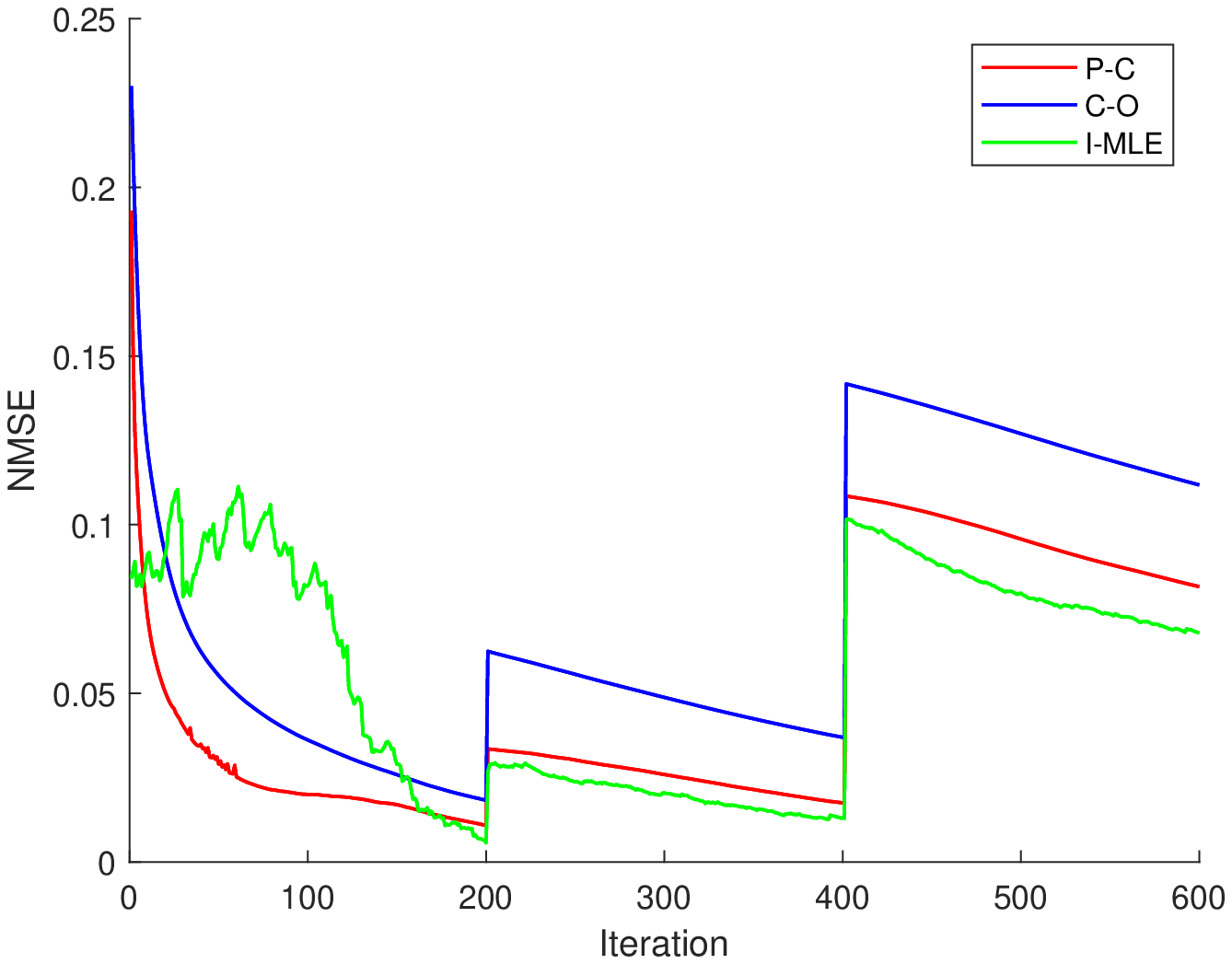}%
\caption{$N=8$, $P=1$, $C=1$}%
\label{sfig}%
\end{subfigure}%
\begin{subfigure}{0.33\textwidth}
\centering

\includegraphics[width=.95\textwidth]{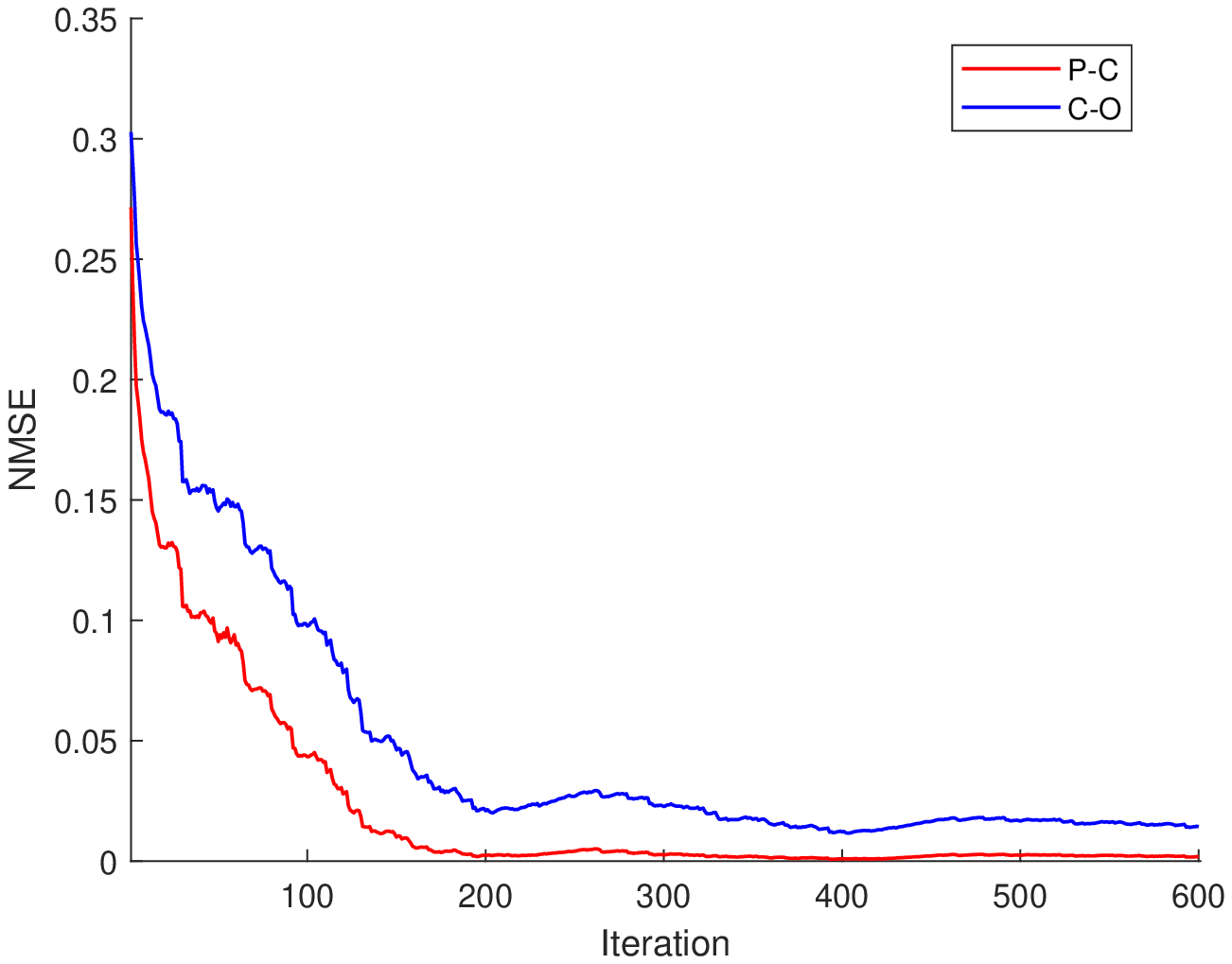}
\caption{$N=8$, $P=1$, $C=1$}%
\label{cfig}%
\end{subfigure}%
\begin{subfigure}{0.33\textwidth}
\centering
\includegraphics[width=.95\textwidth]{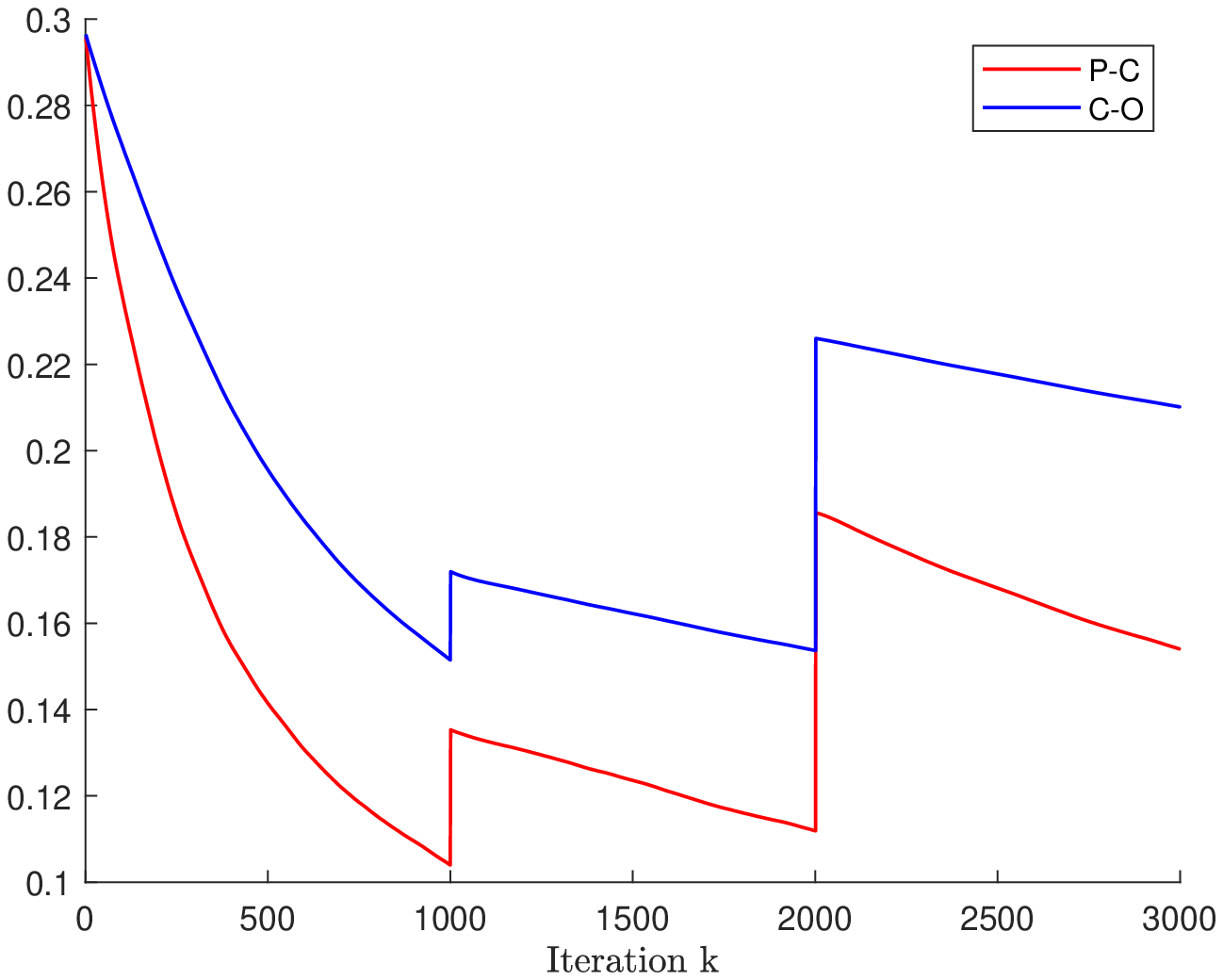}%
%\vspace{.7cm}
\caption{$N=128$, $P=1$, $C=1$ }%
\label{rfig}%
\end{subfigure}%
\caption{(a) NMSE of the (P-C), (C-O), (I-MLE) solutions w.r.t. the (B-MLE) solution; (b) NMSE of the (P-C), (C-O) solutions w.r.t. the (I-MLE) solution; (c)  NMSE of the (P-C), (C-O) solutions w.r.t. the true underlying graph topology.}
\label{fig:NMSE}\vspace{-5mm}
\end{figure*}

%For the sanity check of the algorithm, w
We consider a simple graph of $N=8$ nodes undergoing two triggering events. For the graph and data generation, we adopt the following procedure: first, we generate a positive definite matrix representing the precision matrix $\bbS_0$ at time $t=0$, i.e., the starting graph topology, that remains constant for the first $200$ time instants. Then, we perturb the graph topology obtaining a new GSO, that, again,  remains constant for $200$ time indices. We repeat this again, obtaining a third GSO.
Each time, we perturb the graph by randomly selecting one node, and increasing the weight of its adjacent edges with $20\%$. Finally, we draw $T=600$ graph signals following $\bbx_t \sim \ccalN(\mathbf{0},\mathbf{\Sigma}_t)$, where $\mathbf{\Sigma}_t= \bbS_t^{-1}$, for $t=\{1, \ldots, T\}$.

We assume that each graph signal arrives in an online fashion one at the time. In this way, our cost function changes at every new sample that arrives. However, the distribution of the data, reflected in the precision matrix $\bbS_t$, changes only every $200$ samples.

\smallskip
\noindent \textbf{Analysis.} We are interested in tracking: \textit{i)} the optimal batch MLE (\textbf{B-MLE}) solution that solves the standard GGM \eqref{eq:ggm} for the three stationary intervals of $200$ graph signals (should be close to the generative precision matrix $\bbS_t$); and \textit{ii)} the instantaneous MLE (\textbf{I-MLE}) solution that solves \eqref{eq:graphical-lasso} and makes use of the streaming graph signals up to a particular time instant. We denote the B-MLE and I-MLE solutions as $\bbS_\text{B}$ and  $\bbS_\text{I}$, respectively. Hence, we have $3$ B-MLE and $600$ I-MLE solutions, equal to the number of stationary intervals and time instants, respectively. If, at the end of each stationary period, the empirical covariance matrices coincide, then $\bbS_\text{I}$ should coincide with $\bbS_\text{B}$. However, since for the I-MLE solution we use a forgetting factor, they will not.

To assess the validity of our approach, we compare our prediction-correction (\textbf{P-C}) solution, and the correction-only (\textbf{C-O}) solution with respect to : \textit{i)} the I-MLE solution $\bbS_\text{I}$ for each time instant $t \in \{1, \ldots, T\}$; \textit{ii)} the B-MLE solution $\bbS_\text{B}$ for the three stationary intervals. We also compare how $\bbS_\text{I}$ deviates from $\bbS_\text{B}$ at each time instant. We evaluate the performance of the algorithm by means of the normalized MSE (NMSE), computed as:

\begin{equation}
    \label{eq:nmse}
    \text{NMSE}= \frac{\|\hat{\bbS} - \bbS_{\text{MLE}} \|_F^2}{\|\bbS_\text{MLE}\|_F^2}
\end{equation}
where $\hat{\bbS}$ is either the P-C or the C-O solution (or $\bbS_\text{I}$ when comparing to $\bbS_\text{B}$ ), and $\bbS_{\text{MLE}}$ is either $\bbS_\text{I}$ or $\bbS_\text{B}$.

\noindent\textbf{Results.} Fig.~\ref{sfig} shows the NMSE of the P-C solution (for $P=1$,  $C=1$, $\alpha=\beta=1e-3$), the C-O solution ($C=1$, $\alpha=\beta=1e-3$) and I-MLE solution $\bbS_{\text{I}}$ with respect to the optimal B-MLE $\bbS_{\text{B}}$, for a forgetting factor of $\gamma=0.97$. The triggering effect is visible for $t=\{200, 400\}$, where the NMSE has a sharp increase. All the three solutions show a convergence behavior to the B-MLE $\bbS_{\text{B}}$ and we especially observe how the prediction step improves the performances w.r.t. a correction-only algorithm. This is also visible in Fig.~\ref{cfig}, that shows the NMSE of the P-C and C-O solution w.r.t. the I-MLE solution for each time instant. Here we see also how at $t=\{200,400\}$, the iterative solutions move apart from the I-MLE solution, and then follow it again, probably due to the forgetting factor.
We can conclude, based on these two figures, that the algorithm naturally enforces a similarity of the solutions at each iteration, differently from the MLE solution. In other words, the algorithm adds a regularization to the problem without any regularizer in the cost function. This is achieved with an \textit{early stopping} of the descent steps, governed by the parameters $P$ and $C$. 

A more challenging scenario is considered in Fig.~\ref{rfig}, where the algorithm runs on a graph of $N=128$ nodes, and follows the same perturbation model as the previous one, yet this time with a $50\%$ dilation. Due to the high-computational cost involved for the MLE solutions, we show the NMSE of the iterative solutions with respect to the true underlying matrix $\bbS_t$, where we can still see the trajectory tracking of the algorithm.

Experimentally, we also noticed (not shown) how we can tune the behavior of the algorithm between following the I-MLE solution ( increasing $C$) or the B-MLE solution (using a low $P$ and $C$). We also observed how, for a low value of $C$, the value of $P$ has a  an impact on the performance, especially w.r.t. the B-MLE solution.

\vspace{-2mm}
\section{Conclusion}
\vspace{-1mm}
We proposed an online algorithm operating in non-stationary environments to learn a dynamic graph topology from observed data. The proposed approach, built on the prediction-correction framework, does not require the knowledge of the time instants in which the topology changes. It further implicitly regularizes the problem due to the early-stopping behavior in the iterative process, enforcing similarity between solution iterates. Because the algorithmic formulation is developed without explicitly specifying the cost function, it can be used to extend a variety of (static) topology identification algorithms. We detailed its application to the Gaussian graphical model (GGM) problem, where numerical results show the tracking capabilities of the algorithm with respect to the optimal solution(s).
As future works, we will validate our algorithm for different perturbation models and different cost functions, as well as on real data sets. Building on our recent work \cite{natali2020topology}, we also envision to extend it to cases where the sparsity pattern of the graph is assumed to be known, yet not the importance of the edge weights.
% \vfill\pagebreak

\bibliographystyle{IEEEbib}
\bibliography{refs}

\end{document}